\documentclass[twocolumn,showpacs,preprintnumbers,nofootinbib,amsmath,amssymb]{revtex4}
\usepackage{epsfig}
\usepackage{graphicx}
\usepackage{dcolumn}
\usepackage{bm}

\begin{document}

\title{Solving three-body scattering problem in the momentum lattice representation}

 \author{V.N. Pomerantsev}
 \email{pomeran@nucl-th.sinp.msu.ru}
\author{V.I. Kukulin}
 \email{kukulin@nucl-th.sinp.msu.ru}
\author{O.A. Rubtsova}%
 \email{rubtsova-olga@yandex.ru}
\affiliation{%
Institute of Nuclear Physics, Moscow State University, 119992  Moscow, Russia}

\date{\today}

\begin{abstract}
A brief description of the  novel approach towards  solving few-body
scattering problems in a finite-dimensional functional space of the
$L_2$-type is presented. The method is based on the  complete
few-body continuum discretization in the basis of stationary wave
packets. This basis, being transformed to the momentum
representation, leads to the cell-lattice-like discretization of the
momentum space. So the initial scattering problem can be formulated
on the multi-dimensional momentum lattice which makes it possible to
reduce the solution of any scattering problem above the breakup
threshold (where the integral kernels include, in general, some
complicated moving singularities) to convenient simple matrix
equations which can be solved  on the real energy axis. The phase
shifts and inelasticity parameters for the three-body $nd$ elastic
scattering with MT I-III $NN$ potential both below and above the
three-body breakup threshold calculated with the proposed
wave-packet technique
 are in a very good agreement with the
previous accurate benchmark calculation results.
\end{abstract}

\maketitle \underline{1. Motivation of the work}. The strictly proved
integral equations for the solution of few-body  scattering problems
 were developed many years ago by Faddeev and
Yakubosky \cite{Fad,Yak}. After these pioneer works a lot of
investigations in the few-body quantum physics have been done along
these lines for a few last decades. In spite of a great progress in
this field
\cite{lavern,glokle,vlahovic,laz,benchm1,benchm2,fonseca}, the
practical solution of the few-nucleon scattering problems with
realistic $2N$- and $3N$- interactions, especially above the three
body breakup threshold, is still remained rather cumbersome
computational problem which needs an appeal to very powerful
computer resources. Moreover, even till now $4N$ systems above the
three-body threshold can be practically treated  within the
Faddeev--Yakubovsky framework only with simple pairwise local
interactions \cite{fonseca,laz}. The reason is in very laborious
numerical routines in the coordinate space or complicated moving
singularities in kernels of the integral momentum-space equations.

At the same time, several efficient  methods for the approximation
of  few-body continuum wavefunctions  in various $L_2$ bases have
been developed
\cite{reinh,arabi,yak2,papp2,horse,efros,farell,levin,P}. These are
the "moment $T$-matrix method" \cite{reinh}, J-matrix approach
\cite{arabi,yak2,papp2},  "the harmonic oscillator representation"
\cite{horse}, the Lorentz integral transform method \cite{efros},
the continuum-discretized coupled-channel method (CDCC)
\cite{farell,levin,P} etc. However most of them can be used for
special cases of the few-body scattering only, e.g. for the so
called truly few-body scattering  when there are no bound states in
any two-body subsystems \cite{horse}, or for the composite particle
scattering off heavy target when stripping channels can be neglected
\cite{farell,levin,P}. In other cases one   describes  the processes
when the few-body  wavefunctions  in the initial channel are of
bound-state type and $L_2$ basis is used to approximate the
final-state few-body continuum only \cite{efros}, or one treats a
three-body scattering at small energies below the three-body
threshold only \cite{papp2}. So that, with the above $L_2$-type
methods   no  precise calculations  for the basic three-body $n-d$
scattering case above the breakup threshold have been carried out up
to date.\footnote{ The realistic three-nucleon calculations for e.g.
$nd$ (or $pd$) scattering below and above three-body breakup
threshold have been done either with variational method \cite{kiev}
using the Schr{\"o}dinger equation approach or with the Faddeev
equations in the momentum \cite{benchm2} or in the configuration
space \cite{vlahovic}.}

Thus, it would be very convenient to have in our disposal a
sufficiently universal method for general continuum discretization
in different two- and few-body scattering problems (in nuclear,
atomic, hadronic etc. physics), which operating with $L_2$ functions
only and non-singular matrix equations both below and above the
breakup thresholds.

Few years ago the present authors have developed a new approach
to solving few-body scattering problems based on discretization of
continuous spectrum of total Hamiltonian~\cite{K1,K2,K3,K4,Moro}.
The method uses the stationary wave packets, i.e. are $L_2$-type functions,
instead of the exact scattering wave functions.
In these works an original
wave-packet formalism have been developed which allowed to construct
finite-dimensional (f.-d.) approximations for basic scattering-theory operators
and find the scattering observables using such approximations. The approach have
recently been tested for the elastic scattering and breakup of a
composite projectiles scattered off heavy targets (with neglecting
the stripping processes), and a perfect agreement with the
CDCC method results have been found \cite{K4,Moro}. In the present
paper we extend our wave-packet approach much further, towards
solving a general three-body scattering problem on the base of the
projected Faddeev equations and illustrate
 it on example of quartet and
doublet $n-d$ scattering below and above three-nucleon breakup
threshold\footnote{As far as the present authors are aware this is
the first precise fully $L_2$ approximated solution for the Faddeev
equation above the three-body threshold.}.


\underline{2. Formulation of the approach}. Here we describe the
three-body wave-packet discretization procedure for the elastic
$n-d$ scattering. The elastic amplitude $X$ for the quartet case can
be found  from a single integral Faddeev equation \cite{schmid}
\begin{equation}
\label{eqx} X=-Pv_1-Pv_1G_1X,
\end{equation}
where $v_1$ is the triplet $NN$ interaction potential,
$G_1=(E-H_1)^{-1}$ is the three-body resolvent of the channel
Hamiltonian $H_1=H_0+v_1$ and $P$ is the permutation operator.


Let us introduce some finite basis
$\{|S_i\rangle\}_{i=1}^{\cal N}$ such that the projector
$\Gamma_{\cal N}$ onto the f.-d. basis subspace can (in some not rigorous
sense) approximate the unit operator $
 \Gamma_{\cal N}=\sum_{i=1}^{\cal N}|S_i\rangle\langle S_i| \to 1$.
Then one can define the ${\cal N}$-dimensional approximation for any
operator $A$  as its projection onto the respective basis subspace:
$\Gamma_{\cal N} A \Gamma_{\cal N} =\sum_{ij}|S_i\rangle
A_{ij}\langle S_j|$, with corresponding matrix elements
$A_{ij}\equiv \langle S_i|A|S_j\rangle$. Using such matrix
approximations for the scattering operators,
  the initial integral
equation can be reduced to the respective matrix equation. We will
denote the matrices of projected operators with  corresponding bold
letters. So,  one finds the
matrix equation instead of the integral equation (\ref{eqx}):
\begin{equation}
\label{lattice_eqx} {\bf X}=-{\bf P}{\bf v}_1-{\bf P}{\bf v}_1{\bf
G}_1{\bf X}.
\end{equation}
Thus, it looks like  it would be possible to find some approximate
solutions of the initial integral equation using some appropriate
$L_2$ bases through the simple matrix algebra. However, not everyone
$L_2$ basis suits for this purpose.  The integral kernel of the
Faddeev equation includes the fixed-pole singularities and also  the
complicated moving singularities above the three-body breakup
threshold. Just these singularities correspond to the proper
boundary conditions in coordinate space and  provide the correct
physical solution of the Faddeev equations, but  a construction of
the appropriate basis for the projection of such kernels is highly
non-trivial problem. Another key problem here is a calculation of
matrix elements for $v_1$, $P$ and especially for $G_1$ operators in
the chosen basis, that can appear, in general, rather cumbersome
task. In this point the appropriate stationary wave-packet basis
helps to overcome all the above difficulties and formulate the
matrix equations whose solutions can really approximate the
solutions of the initial integral equations. Such a basis will be
demonstrated to provide
 a convenient analytical diagonal
representation for the three-body channel resolvent matrix ${\bf
G}_1$ and, on the other hand, this basis covers a sufficiently wide functional
$L_2$ space to provide well converged results.

 The
channel Hamiltonian $H_1$ is  the direct sum of
two subHamiltonians corresponding to the system motion along two
independent Jacoby coordinates:
 \begin{equation}
H_1\equiv h_1\oplus h_0,
\end{equation}
where subHamiltonian $h_1$ defines the $NN$ subsystem motion
(including interaction $v_1$) and subHamiltonian $h_0$
 corresponds to the free motion of the third nucleon relatively the $NN$
 subsystem center of mass. Now let's
 define two sets of momentum bins $[p_{m-1},p_m]_{m=1}^M$ and
 $[q_{n-1},q_n]_{n=1}^N$ corresponding to the continuum discretizations for  subHamiltonians
 $h_1$ and $h_0$.
 The (two-body) stationary
 wave-packet bases (WPB)  are defined as
 integrals of exact continuum wave functions $|\psi_p\rangle$ and
 $|\psi_{0q}\rangle$  of  subHamiltonians $h_1$ and $h_0$ over the respective
 momentum bins:
 \begin{equation}
 |z_m\rangle=\frac1{\sqrt{b_m}}\int_{p_{m-1}}^{p_m}{\rm
 d}p|\psi_p\rangle,\quad
|y_n\rangle=\frac1{\sqrt{d_n}}\int_{q_{n-1}}^{p_n}{\rm
 d}q|\psi_{0q}\rangle,
 \end{equation}
 where $b_{m}\equiv p_{m}-p_{m-1}$ and $d_{n}\equiv q_{n}-q_{n-1}$
 are bin widths. Now, the  three-body WPB states $|S_{ij}\rangle$ are defined
just as products of two-body wave-packet states $|z_i\rangle$
(including the bound state wave function $|z_0\rangle$) and
$|y_j\rangle$:
 \begin{equation}
|S_{ij}\rangle \equiv|z_{i},y_{j}\rangle.
\end{equation}
We omit here partial wave indices for the sake of simplicity.
 The
properties of the wave-packet sets $|z_i\rangle$ and $|y_j\rangle$
have been investigated in detail in refs.~\cite{K1,K2,K3,K4}. In
particular, the very useful property of such a packet basis is that
the matrices for the projected resolvents of  subHamiltonians
$g_1(E)=(E+{\rm i}0-h_1)^{-1}$ and $g_0=(E+{\rm i}0-h_0)^{-1}$ are
diagonal and  defined by a simple analytical expressions depending
only on the spectrum discretization parameters. This property of WPB
allows us to construct the f.-d. analytical diagonal representation
for the channel resolvent $G_1(E)$, which is a convolution  of
two-body resolvents $g_1$ and $g_0$.

Indeed,  the exact three-body channel resolvent can
be written \cite{K2} as a sum of two terms $G_1(E)=G_1^{\rm
BC}(E)+G_2^{\rm CC}$, where  the bound-continuum part  has the form:
\begin{equation}
G_1^{\rm BC}(E)=\int_{0}^\infty {\rm d}q
\frac{|z_0,\psi_{0q}\rangle\langle z_0,\psi_{0q}|}{E+{\rm
i}0-\epsilon_0-\frac{3q^2}{4m}},
\end{equation}
and $\epsilon_0$ is the binding energy for the (single) $NN$ bound
state.   The continuum-continuum part takes the form:
\begin{equation}
G_1^{\rm CC}(E)=\int_{0}^\infty {\rm d}p \int_{0}^\infty {\rm d}q
\frac{|\psi_p,\psi_{0q}\rangle\langle \psi_p,\psi_{0q}|}{E+{\rm
i}0-\frac{p^2}{m}-\frac{3q^2}{4m}}.
\end{equation}
Now let's construct a projection of the exact channel resolvent onto
the three-body WPB. The following analytical formulas for the
diagonal f.-d. projection of $G_1$ can then be found:
\begin{eqnarray}
 {\bf G}_1^{\rm BC}=\sum_{j} G^{BC}_{0j}
|z_0,y_j\rangle\langle z_0,y_j|,\nonumber \\ {\bf G}_1^{\rm
CC}=\sum_{i\neq0,j} G^{CC}_{ij} |z_i,y_j\rangle\langle z_i,y_j|,
\label{dg1}
\end{eqnarray}
where the  matrix elements $G_{oj}^{\rm BC}$ and $G_{ij}^{\rm CC}$
in eq. (\ref{dg1}) are defined  as integrals over the respective
momentum bins:
\[
G^{BC}_{0j}=\frac1{d_j}\int_{q_{j-1}}^{q_j}\frac{{\rm d}q }{E+{\rm
i}0-\epsilon_0-\frac{3q^2}{4m}}, \eqno(\ref{dg1}a)
\]
\[
G^{CC}_{ij}=\frac1{d_id_j}\int_{p_{i-1}}^{p_i}
\int_{q_{j-1}}^{q_j}\frac{{\rm d}p{\rm d}q }{E+{\rm
i}0-\frac{p^2}{m}-\frac{3q^2}{4m}}. \eqno(\ref{dg1}b)
\]
These matrix elements  depend, in general, on the spectrum partition
parameters (i.e. $p_i$ and $q_j$ values). However we found  when the
wave-packet expansions of the three-body amplitude is convergent the
final result turns out to be {\em independent} upon the particular
spectral partition parameters. The integrals in eqs.(\ref{dg1}a) and
(\ref{dg1}b) are analytically tractable that gives a simple
analytical f.-d. representation for the three-body channel resolvent
$G_1$. Such an analytical representation  is main feature of the
wave-packet approach since it allows to simplify solution of the
general three-body scattering problem  drastically.

Now the key question arises: how  to construct practically the above
wave-packet basis. The free packets $|y_{n}\rangle$ corresponding to
the free motion of third nucleon relatively the $NN$ subsystem c.m. take in the
momentum representation the form of simple step-like functions:
\begin{equation}
\langle q|y_j\rangle
=\frac{\theta(q-q_{j-1})-\theta(q-q_{j})}{\sqrt{d_j}},\quad
\end{equation}
where $\theta (x)$ is the Heaviside function.

 The scattering wave packets
$|z_{i}\rangle$ describing the scattering in the $NN$ two-body
subsystem can be very well approximated by pseudostates
$|\tilde{z}_{i}\rangle$ obtained from the diagonalization of the
subHamiltonian $h_1$ in some appropriate $L_2$ basis \cite{K1}. In
the present work we use for this diagonalization a {free}
wave-packet basis $|x_k\rangle$ corresponding to the free $NN$
motion. Thus, we solve the two-body variational problem directly on
the free WPB  and as a result   obtain a set of variational
functions
\begin{equation}
|\tilde{z}_i\rangle=\sum_{k=0}^{M} O_{ik}|x_k\rangle, \qquad
i=0,\dots, M, \label{exp_z}
\end{equation}
the first of which (for the  problem in question) is the
wavefunction of the bound state (deuteron) and other ones are very
good approximations for the exact scattering packets.
\begin{figure}
\begin{center}
\epsfig{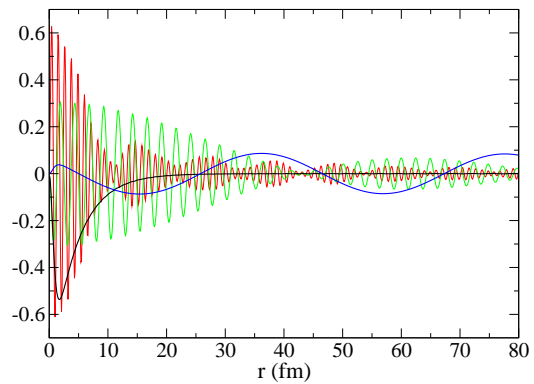}
\end{center}
\caption{\label{fig1} (Color online) Bound state (solid curve) and several
interaction wave packets (different dashed lines) for the MT III
potential constructed from the free momentum packets in the
coordinate space.}
\end{figure}

In the Fig.~1 the coordinate-space behavior  of some from the first
50 variational functions (including the deuteron)  is shown. It is
clear that the free packet basis allows to approximate the
respective scattering wave functions up to a very far asymptotic
region (in the Figure the functions $\tilde{z}_i(r)$ are given at
$r<80$~fm, but actually they  coincide with  exact scattering wave
packets up to $r \sim$ 1000 fm). This long-range behavior of the basis
functions plays a crucial role in the three-body scattering,
especially above the breakup threshold, because it provides a proper
overlapping between basis functions in different Jacoby-coordinate
sets. It should be mentioned that in our previous calculations
\cite{K2} we used Gaussian basis to approximate the interaction
packets in the whole space and so the wrong long-range behavior of
the  basis functions did not allow us to obtain well converged
results above the break-up threshold (while for the smaller energies
the Gaussian approximation works well)\footnote{It should be noted
the free packet basis (i.e. the step-like functions in the momentum
space) is, of course, not optimal for calculation of bound states.
E.g. in our case only 20 Gaussian functions are necessary to obtain
$E_b=-2.225$~MeV and ca. 100 step-like functions are required to
reach the same precise bound energy value. However very good
approximation of scattering wave functions in two-body subsystems is
the decisive factor here.}.

Besides, the momentum lattice basis is very convenient to find the
matrix elements of  the permutation operator $P$. Using
approximation (\ref{exp_z}) for the scattering packets
$|z_{i}\rangle$, these matrix elements can be expressed in the form:
 \begin{equation}
P_{ij,i'j'}=\langle z_{i}y_{j}|P|z_{i'}y_{j'}\rangle \approx
\sum_{kk'}O_{ik}O_{i'k'}^*P^0_{kj,k'j'},
\end{equation}
where $P^0_{kj,k'j'}\equiv\langle x_ky_j|P|x_{k'}y_{j'}\rangle$  is
the permutation  matrix elements taken on the two-dimensional free
wave packets (for a two-dimensional lattice). Using hyperspherical
momentum coordinates, the calculation of $P^0_{kj,k'j'}$ can be
reduced to a
 one-dimensional numerical integration over hypermomentum
$p^2+q^2$. The  technique of this calculation will be given in
detail elsewhere. It should be stressed here that this is one of the
key points for the whole our approach. In fact, to solve
two-dimensional Faddeev equations by conventional
methods~\cite{glokle} one needs (due to the appearance of
permutation operator $P$ in the integral kernel) to use a very
time-consuming two- and three-dimensional interpolations (many
thousands or even millions of such interpolations) at each iteration
step to find the solution in the initial Jacoby set from the
``rotated'' (by the permutation $P$) Jacoby sets. So, such numerous
multi-dimensional interpolations at each step take a big portion of
computational time in practical solutions of three-body integral
equations. When solving the four-body Yakubovsky equations the
dimension and number of the above each step interpolations gets even
higher. Thus, the wave-packet approach allows to avoid such
multi-dimensional interpolations.

After solving the matrix eq. (\ref{lattice_eqx}),  the on-shell
elastic amplitude $A_{\rm el}(E)$ in wave-packet approximation  can be
found as a diagonal (on-shell) matrix element  of $X$-matrix (which is a
solution of the matrix equation (\ref{lattice_eqx})):
\begin{equation}
\label{el}  A_{\rm el}(E)\approx
\frac{2m}{3q_{0}}\frac{X_{0n_0,0n_0}}{d_{n_0}},
\end{equation}
where index $n_0$ being denote the singular $q$-bin  to which the
on-shell momentum $q_0=\sqrt{\frac{4}{3}m(E-\epsilon_0)}$ belongs:
 $q_0\in (q_{n_0-1},q_{n_0})$. Let's notice that in order to find the
elastic amplitude according to eq. (\ref{el}), one needs to solve
 one linear equation only just for  one column
$X_{mn,0n_0}$, but not to do a full matrix inversion in
eq.~(\ref{lattice_eqx}).

\underline{3. Numerical results}. To illustrate the accuracy and
effectiveness of the proposed wave-packet technique we calculated
the real phase shifts and inelasticity parameters for the three-body
elastic $n-d$ scattering in the quartet and doublet $S$-wave
channels with the model Malfliet-Tjon $NN$ potential MT-III.
 The results of these calculations  are
 shown in Figs.~2 and 3 for the spin-quartet and in Figs.~4 and 5 for the
 spin-doublet channels respectively.
 \begin{figure}[hpt]
\begin{center}
\epsfig{file=fig2.eps,width=0.8\columnwidth}
\end{center}
\caption{(Color online) The energy dependence of the real phase
shift for $S$-wave quartet $nd$ scattering calculated by means of the
momentum-packet discretized Faddeev equation at different dimensions
$M\times N$ of the lattice basis: $100\times 100$ (dashed curve),
$200\times200$ (solid curve). Results of the direct Faddeev equation
solution from ref.\cite{benchm1,benchm2} are marked as
$\blacktriangle$.} \label{fig:phase_qu}
\end{figure}

\begin{figure}[hpt]
\begin{center}
\epsfig{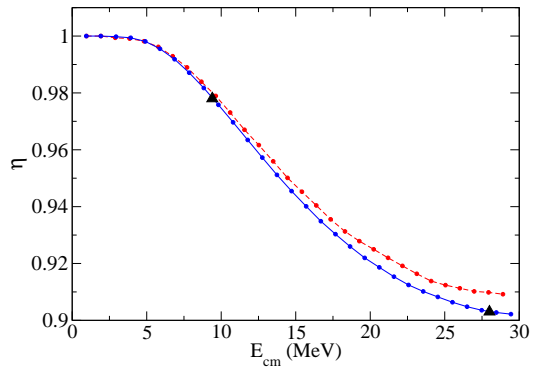}
\end{center}
\caption{(Color online) The same as in Fig.~\ref{fig:phase_qu} but for the
inelasticity parameter $\eta$.} \label{fig:inel_qu}
\end{figure}
In the case of doublet
 scattering  one has the system of two matrix equations instead of one matrix
  equation (\ref{lattice_eqx})
  where two
 amplitudes correspond to two possible spin states (triplet and singlet) of the $NN$
 subsystem. To check the accuracy of the method we have compared our
 results with the previous benchmark calculation results from ref.~\cite{benchm1}
 (below the deuteron breakup threshold)  and ref.~\cite{benchm2} (above the deuteron breakup threshold)
   The exact parameters of the $NN$
 potential are taken from ref.~\cite{benchm2}.

 \begin{figure}[hpt]
\begin{center}
\epsfig{file=fig4.eps,width=0.8\columnwidth}
\end{center}
\caption{(Color online) The energy dependence of the real phase shift for the 
$S$-wave
doublet $nd$ scattering calculated by means of momentum-packet
discretized Faddeev equation at different dimensions $M\times N$ of
the lattice basis: $(50+50)\times50$ (dashed curve), $(80+80)\times80$ (dotted
curve) and $(100+100)\times100$ (solid curve). Results of the direct
Faddeev equation solution from ref.\cite{benchm1,benchm2} are marked as
$\triangle$. } \label{fig:phase_du}
\end{figure}

As it is seen from the Figs.~2-5 the wave-packet discretization
technique  for the three-body continuum  works successfully for the
general three-body scattering problem both below and above breakup
threshold. Thus,  at the first time we have solved the three-body
scattering problem above the breakup threshold using only f.-d.
approximation  of the $L_2$-type for the Faddeev kernel. Just the
use of the momentum-lattice basis  allowed us to achieve a good
convergence and accuracy  on this way. It is interesting to remark
that although the quartet case seems  to be simpler from the first
glance (one equation and the inelasticity is  less than that  in the
doublet case) it turns out to be more difficult numerical problem (the larger basis is
needed for convergence) for the wave
packet discretization approach.

\begin{figure}[hpt]
\begin{center}
\epsfig{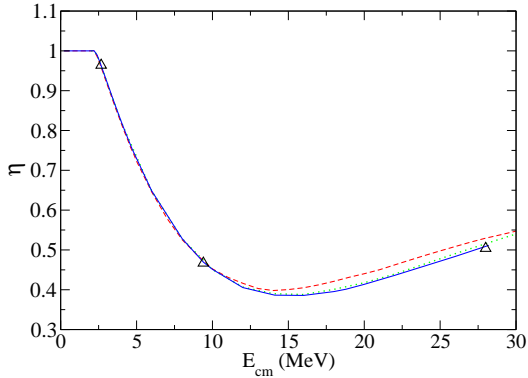}
\end{center}
\caption{(Color online) The same as in Fig.~\ref{fig:phase_du} but for the
inelasticity parameter $\eta$.} \label{fig:inel_du}
\end{figure}

\underline{4. Conclusion.} Let's briefly outline here the most
important points of this study. For the first time the three-body
scattering problem in the Faddeev framework above the breakup
threshold has been successfully solved in the three-body $L_2$ basis
representation using the lattice approximation scheme (which is the
technique of the three-body continuum discretization).  The success
and  advantages   of the  lattice approach  are related to the
following distinctive features.

(i) The explicit analytical f.-d. approximation for the three-body
channel resolvent $G_1$  allows to reduce initial integral Faddeev
equation to the matrix one that can be solved directly on the real
energy axis.

(ii) The  scattering wave packets (corresponding to the $NN$
interaction) can be approximated by pseudostates of the two-body $NN$
subHamiltonian matrix in the free wave-packet basis, which allows to
avoid calculation of the two-body $t$-matrix and obtain
 explicitly  matrix elements of the permutation  operator $P$ that
includes overlapping between  wave-packet basis states of the
different three-body channel Hamiltonians.

 (iii) This convenient closed form for the matrix elements of the
permutation operator $P$ in the WPB also makes it possible to avoid
completely very time-consuming multi-dimensional interpolations of
the iterated kernels which are usually assist in conventional
techniques of  the   Faddeev equation numerical integration in the
momentum space.

(iv) The very
 long-range type of the wave-packet  functions (non-vanishing
 at distances $\sim$ 1000 fm) allows to
 approximate properly the overlapping between basis states
 in different Jacoby coordinate sets. This leads to the proper
 asymptotic behavior of the solutions along different Jacoby
 coordinates, which couldn't be provided by means of conventionally used
short-range type $L_2$ bases.

 Besides, this
long-range behavior of the wave-packet basis functions  looks also
very promising for the proper incorporation of the long-range
Coulomb interaction in the treatment of the few-nucleon scattering.
Our further investigations are pointed at this purpose.

{\bf Acknowledgements.} The present authors appreciate greatly
  partial financial supports from the RFBR grant 07-02-00609,
the joint RFBR--DFG grant 08-02-91959 and the Russian President
grant for young scientists  MK-202.2008.2.

\end{document}